\documentclass[nameyear]{elsart}
\usepackage{epsfig}

\newcommand{\be}{\begin{eqnarray}}
\newcommand{\ee}{\end{eqnarray}}
\def\lsim{\mathrel{\rlap{\lower3pt\hbox{\hskip1pt$\sim$}}
    \raise1pt\hbox{$<$}}} 
\def\gsim{\mathrel{\rlap{\lower3pt\hbox{\hskip1pt$\sim$}}
    \raise1pt\hbox{$>$}}} 
\newcommand{\msun}{M_\odot}
\newcommand{\rsun}{R_\odot}

\def\bi{\begin{enumerate}}
\def\ei{\end{enumerate}}

\def\astrobjNovaSco{Nova Scorpii 1994}
\def\astrobjV404Cyg{V404 Cyg }
\def\astrobjCygX1{Cyg X-1}

\begin{document}

\runauthor{Brown, Lee, and Tauris}

\begin{frontmatter}
\title{Formation and Evolution of Black Hole X-ray Transient Systems}
\author[suny]{G.E. Brown\thanksref{geb}}
\author[suny,kias]{C.-H. Lee\thanksref{chl}}
\author[nordita]{T.M. Tauris\thanksref{tt}}
\address[suny]{Department of Physics \& Astronomy,
        State University of New York,\\
        Stony Brook, New York 11794, USA}
\address[kias]{School of Physics,
       Korea Institute for Advanced Study, Seoul 130-012, Korea}
\address[nordita]{Nordic Institute for Theoretical Physics (NORDITA), 
Blegdamsvej 17, DK-2100 Copenhagen \O, Denmark}
\thanks[geb]{E-mail: popenoe@nuclear.physics.sunysb.edu}
\thanks[chl]{E-mail: chlee@kias.re.kr}
\thanks[tt]{E-mail: tauris@nordita.dk}

%
\begin{abstract}
We study the formation of low-mass black hole X-ray 
binaries with main sequence companions that have formed 
through case C mass transfer (mass transfer following the helium
core burning phase of the black hole progenitor). We identify 
these objects with the observed soft X-ray transients.
Although this scenario requires a set of fine tuned conditions, we are
able to produce a current Galactic population of $\sim$2000 objects,
in agreement with estimates based on observations. 
  
The narrow interval in initial separations leading to case~{C} mass
transfer, combined with the allowed narrow range of separations after 
the common envelope evolution, constrains 
the common envelope efficiency in this scenario: 
$\lambda \alpha_{ce} \approx 0.2-0.5$.
\end{abstract} 


\end{frontmatter}

%

\newpage
\section{Introduction}

%
%

In the literature there have been severe difficulties in evolving a 
sufficient number of black hole soft X-ray transient sources (SXTs)
Ergma \& van den Heuvel (1998), Portegies Zwart, Verbunt, \& Ergma (1997).
For example, Portegies Zwart et al. obtain a birth rate of
$9.6\times 10^{-9}$ yr$^{-1}$ for the SXTs, to be compared with one
of $2.2\times 10^{-6}$ yr$^{-1}$ for the binaries with a neutron star,
whereas they remark that observations indicate equal formation rates
for these two types of binaries. Their large discrepancy arises from
their limit ZAMS mass $40\msun$ for black hole formation. 

In fact, however, for the evolution of the SXTs, a ZAMS mass more like
$\sim 20\msun$ should be chosen as limit for high-mass black hole formation.
This is the limit for single stars (Brown et al. 2001).
If the massive black hole progenitor in the SXTs can complete He core burning
before it is removed in common envelope evolution (Case C mass transfer)
then it will evolve like a single star since the remaining $\sim 10^4$ yrs
of its lifetime is too short for He wind losses to effect its evolution.
Indeed, Ergma \& van den Heuvel (1998) realized that ``$M_{BH}$ should not
be larger than $20$ to $25\msun$" in order to have sufficient SXTs, but 
they did not realize that the more massive stars in a binary evolve in
a different way from a single massive star.

If the H envelope of the more massive star in a binary is removed by early
Roche Lobe overflow, Case A or Case B, before He core burning is completed,
the resulting ``naked" He core will blow away, leaving an Fe core too low
in mass to evolve into a high-mass black hole (Brown et al. 2001).


The common envelope evolution of Portegies Zwart et al. (1997) for a
$20\msun$ turns out to be useful for us, because this will be the same,
regardless of whether the resulting He star goes into a neutron star as
they believed, or into a high-mass black hole. It turns out that this
evolution can only be carried out in Case C, so the He star is clothed
during He core burning, appropriate for our high mass black hole evolution.

Given the Schaller et al. (1992) evolution, we find that Case C evolution
does not work for stars more massive than $\sim 22\msun$, in agreement
with Portegies Zwart et al.(1997). However, we need the more massive
black hole progenitors to evolve the transient sources with subgiant
companions, as we shall discuss.

In this note we consider the evolution of those low-mass black hole X-ray
binaries with main sequence companions which are
formed through case~{C} mass transfer followed by common 
envelope and spiral-in evolution. We estimate their expected number
currently present in our Galaxy. We therefore discuss the following.

\begin{enumerate}
\item 
The initial separation interval of binaries 
that go through case C mass transfer and the ZAMS mass 
interval of stars being able to produce high-mass black holes (Sect.~2).
\item The initial masses of the black hole companions (the donors 
  in the present-day X-ray binaries) and the allowed orbital periods 
  after the common envelope required for the systems to become
  observable X-ray binaries within the age of our Galaxy before
  the donors leave the main-sequence (Sect.~3).
\item The resulting common envelope efficiency that can be estimated
  combining points 1 and 2 (Sect.~4).
\item The evolution of the mass-transfer rate in the X-ray transient
  phase and the lifetimes of the systems (Sect.~5).
\item The total number of systems expected in our Galaxy at 
  present (Sect.~6).
\end{enumerate}

Following the discussion of the above 5 points, we briefly discuss
the black hole systems with (sub)giant companions Nova Scorpii (GRO J1655$-$40)
with a $2.4\msun$ F-star companion and a $6.3\msun$ black hole and V4641 Sgr
(XTE J1819$-$254) with a $6.5\msun$ B-star companion and a $9.6\msun$ black
hole. At least in the former case there has been substantial mass loss in
the black hole formation, $\gsim 5\msun$. Thus, a He core of at least
$11\msun$ is needed for the black hole progenitor, or a ZAMS $35\msun$ star.
From this we conclude that the Schaller et al. (1992) wind losses are
too large and that they must be substantially decreased in order to evolve
these two binaries.


\section{Case~{C} mass transfer: Limits on Initial Separations and
         Black Hole Progenitor Masses}
\label{sec-caseC}

In order to have case~{C} mass transfer the radius of the star has
to expand after core helium
burning has stopped. In the calculation of Schaller et al.(1992), 
stars with ZAMS masses $20\msun$ and $25\msun$ satisfy the requirement.
However, during the core helium burning stage, the orbit expands due to
the mass loss:  
\be 
   \frac{a^\prime_{\Delta M}}{a}=\frac{M+M_d}{M-\Delta M +M_d}
\ee
where $a$ is the orbital separations, $M$ is the mass of the black 
hole progenitor, $\Delta M$ is
the mass lost from the black hole progenitor during the core
helium burning, and $M_d$ is the donor star mass.
In order to initiate mass transfer after core helium burning
the star has to expand sufficiently that this widening of the orbit is
compensated for.

\begin{figure}[t!]
\begin{center}
\epsfig{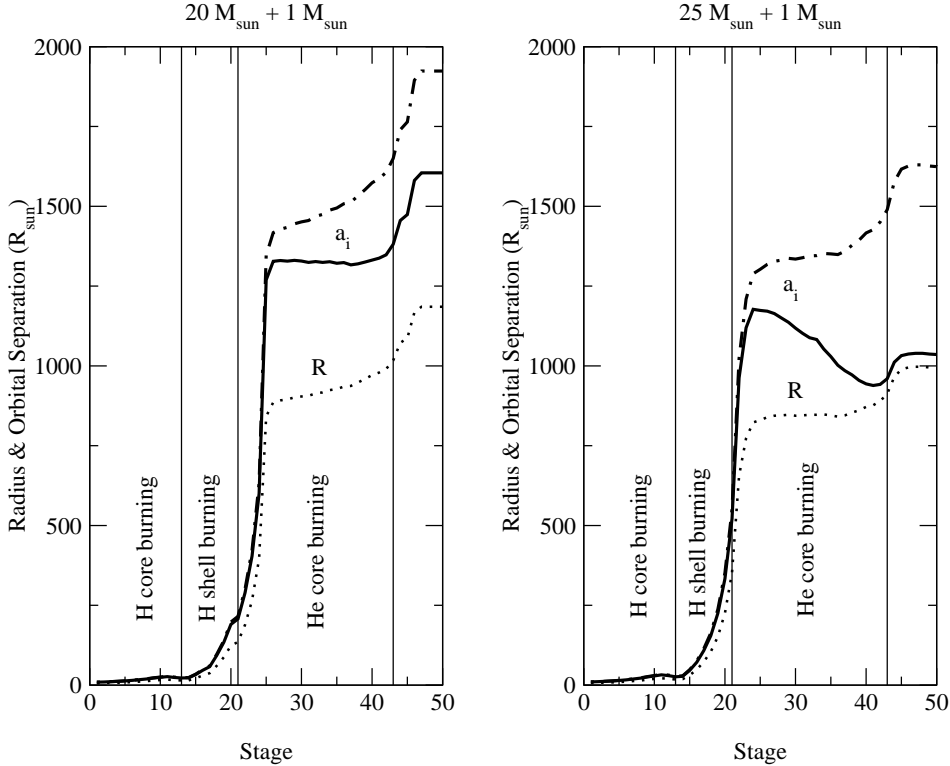}
\end{center}
\caption{Radius of black hole progenitors ($R$) and the
  initial orbital separations ($a_i$) of the progenitors of X-ray transient
  binaries with a $1\msun$ companion. Upper dot dashed curves correspond to
  the orbital separations required to initiate Roche lobe overflow at
  the radius taken for each stage from Schaller et al.~(1992). The 
  solid curves correspond to the required initial
  separations after corrections of the orbit widening due to the 
  wind mass loss. The lower dotted curves correspond to the radius of
  the black hole progenitors taken from Schaller et al.~(1992).
  }
\label{fig-radius}
\end{figure}

In the upper curve of Fig.~\ref{fig-radius} we draw the orbital
separation of the binary required to initiate Roche lobe overflow 
calculated at each stage from the radii given by Schaller et al.~(1992).
For the masses considered the radius of giant star is at 
$\sim 2/3$ of the radial
separation between the two stars.
The solid curve gives the corresponding \emph{initial}
separation between the
massive star and the low-mass main sequence companion after strong wind 
mass loss is switched on. 
Here we assume that the wind mass loss does not affect the stellar 
radius.\footnote{
One of us (T.M.T.) has carried out stellar evolution calculations, using
Eggleton's code, with/without the wind mass losses, and find that the 
radius behaves almost the same.  The minor difference will give rise 
to adjusting our numbers slightly, but the outline
will remain the same.
}
We see that in the case of a $20\msun$ star the
mass transfer can start following He core burning for binaries with an
initial separation in the interval of $1370\rsun < a_{\rm i} <
1605\rsun$. In the case of a $25\msun$ star, however, mass transfer is
only possible up to just after the beginning of core helium burning
(with initial separations around 1170 $\rsun$). Binaries with larger 
initial separations will become too wide during core helium burning so 
even though the star expands slightly after core helium burning, it
will not be able to fills its Roche lobe
(see also Fig. 2 of Portegies Zwart, Verbunt, \& Ergma~1997).

Brown, Weingartner \& Wijers~(1996) found that stars with ZAMS masses
above $19\msun$, evolved by Woosley \& Weaver~(1995), evolve into 
high-mass ($\sim 7\msun$) black holes. 
As outlined in Brown et al.~(2001), this 
requires the special rate of 170 keV barns at energy
$E=300$ keV for the
$^{12}$C$(\alpha,\gamma)^{16}$O reaction used by Woosley \& Weaver. 
Recent experiments including both E1 and E2 components, obtain
$S^{300}_{tot}=(165\pm 50)$ keV barns (Kunz et al. 2001). 
Brown et al. (2001) discuss quantitatively how changes in this rate
change the ZAMS mass at which the calculated Fe core mass increases rapidly
with mass. The fact that the progenitor mass of SN1987A lay in the narrow
interval of $18-20\msun$ appropriate for forming black holes with the
Woosley and Weaver 170 keV barns gives further support to this value.
A ZAMS $18\msun$ star, which is usually taken as progenitor for SN~1987A,
will form a low-mass ($\sim 1.5 - 1.8 \msun$) black hole or a neutron star.

Hence, the mass interval of stars that can go through case~{C} mass 
transfer and collapse into a high-mass black hole is limited between 19 
and 20--25 $\msun$. We shall (arbitrarily) take 22 $\msun$ as an
upper limit. Stellar evolution calculations with lower wind mass losses
might give a higher mass limit.

Using the above derived limits on the mass of the black hole
progenitors and the initial separation, the rate of formation of black
hole transient sources with main sequence companions we can evolve 
(see Section~\ref{pop}) is
nearly an order of magnitude less than in Brown, Lee \& Bethe (1999)
where wind mass losses were ignored.
We realize that our evolution requires a fine-tuned parameter space,
but we shall argue that it all hangs together. The
exact numbers may alter with changes in stellar evolution models.

\section{Limits on Donor Masses and Orbital Periods}
\label{evol_binary}

During the common envelope evolution the low-mass main sequence star
of mass $\sim 1\msun$ will spiral into the envelope of the massive
giant. The orbit will shrink dramatically and the outcome is a close
binary consisting of the core of the massive giant and the 
low-mass star (Paczy\'nski 1976). 
Shortly after the spiral-in and ejection of the envelope of the giant,
the remaining core will collapse into a black hole.
Whether or not this collapse is associated with mass loss 
is still an open question.
The orbit will then shrink further due to the loss of orbital
angular momentum via magnetic braking and gravitational wave 
radiation (e.g. Verbunt 1990) until the low-mass (donor) star
begins to transfer mass to the black hole -- forming an X-ray binary.
Since the wind mass-loss rate of the low-mass donor star is
very low, and its accretion during the short common envelope phase
is negligible, it will have a mass at the onset of the X-ray phase
which is about equal to its initial mass.

The time evolution of the orbital separation is described as
  \be
  \frac{\dot a}{a}=\frac{2 \dot J_{gw}}{J_{orb}}+\frac{2\dot J_{mb}}{J_{orb}}
  -2 {\dot M}_d \left( \frac{M_{BH}-M_{d}}{M_{BH} M_d}\right).
  \label{evo_orb}
  \ee
Here we assume that the mass lost from the companion
star is all accreted onto the black hole.
The orbital angular momentum loss by gravitational wave radiation 
(gw) and magnetic braking (mb) is given by:
   \be
   \frac{\dot J_{gw}}{J_{orb}}
   &=& -\frac{32 G^3}{5 c^5} \frac{M_{BH} M_d (M_{BH}+M_d)}{a^4} \ {\rm s^{-1}}
   \nonumber\\
   \frac{\dot J_{mb}}{J_{orb}}
   &\approx& -0.5\times 10^{-28} b_{mb}
   \frac{IR_d^2}{a^5} \frac{G (M_{BH}+M_d)^2}{M_{BH} M_d} \ {\rm s^{-1}}
   \ee 
    where $R_d$ is the radius of the donor star, $b_{mb}$ is an
   efficiency parameter given below, and $I \approx 0.1 \, M_d
   R_d^2$ is the moment of inertia of the donor star. 
Here we have assumed a magnetic braking law based on observations
of slowly rotating single stars by Skumanich (1972). (However, this
law may be too strong, i.e. overestimating the dependence on the
angular velocity, see e.g. Stepien 1995). We further assumed the
   the magnetic braking efficiency suggested by
   Kalogera \& Webbink (1998),
   \be 
      b_{mb}(M_d) = \left\{
          \begin{array}{lr}
          0, & M_d\leq 0.37\msun \\
          1, & 0.37\msun <M_d<1.03\msun \\
          \exp(-4.15 (M_d-1.03)), & M_d>1.03\msun \\
          \end{array}\right. .
  \label{magbra}
  \ee
   Here the effect of magnetic braking is assumed to be strongest for stars
   with a mass $\approx 1.0 \msun$. 
   Donor stars with
   mass $>1.5 \msun$ do not have a convective envelope on the main 
   sequence and the magnetic braking effects are therefore negligible,
   so the binary evolution is dominated by 
   gravitational radiation until the mass transfer starts.

\begin{figure}[t!]
\begin{center}
\epsfig{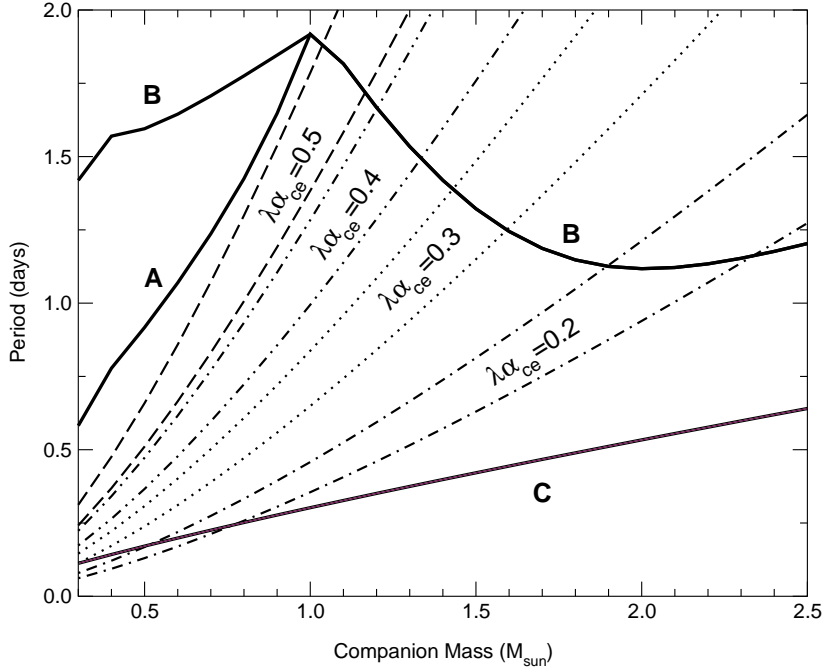}
\end{center}
\caption{
  Limits on the orbital period after the formation of the black hole as
  function of the mass of the low-mass companion. The upper and lower
  boundaries (thick solid lines) of mass and period are set by the
  conditions (see Section~\ref{evol_binary}):
  A) Mass transfer begins within the Hubble time, B) Mass
  transfer begins before the companion evolves off the main sequence,
  C) Mass transfer from the companion begins right after finishing 
  common envelope evolution.
  The expected ranges of periods and companion masses, after the common    
  envelope phase, are plotted for various given efficiencies 
  (see Section~\ref{CE}).
The width of each band for an assumed $\lambda\alpha_{ce}$ is determined
by the initial band of possible Roche Lobe overflows with Case C
mass transfer.
  We took the ZAMS mass of the black hole progenitor to
  be $20\msun$ which corresponds to $M_p\sim 16.8\msun$ in the
  beginning of case~{C} mass transfer, and $M_{He} = M_{BH} = 7\msun$.
  }
\label{fig-possib}
\end{figure}

In Fig.~\ref{fig-possib}, we show the possible ranges of orbital
periods and donor masses immediately after the formation of the black
hole, such that an X-ray binary with a main-sequence donor is formed
within the Hubble time (Kalogera 1999). 
If no mass is lost in the formation of the
black hole, these periods are equal to the periods immediately after
the common envelope as a result of the short time interval before
the collapse of the core.

The upper limit is set by two conditions: i) the binary should start
mass transfer within the Hubble time (line denoted by A), ii) the mass
transfer should start before the donor evolves off the main-sequence
(line denoted by B)  
which also determines the lower limit on
$\lambda\alpha_{ce}=0.2$ (see Section~\ref{CE}). 
The kink in line B around $\sim 1\msun$ comes from the fact that the 
effect of magnetic braking, eq.(\ref{magbra}),
is strongest for stars with a mass $\sim 1\msun$.
For this plot, we took the simple interpolation of
Schaller's results for the evolution of a $1\msun$ star with two
extrapolations for the radius $R$ and the lifetimes of the
main-sequence star 
$t_{\rm end \ ms}$; 
    \be
    R (M_d,t)           &=& R_{\msun} (t) \times M_d^{0.88} \nonumber\\
    t_{\rm end \ ms}(M_d) &=& \left\{ \begin{array}{ll}
    t_{\msun}/M_d^2          & \ \ {\rm for} \ M_d <1\msun \\
    t_{\msun}/M_d^{2.5}      & \ \ {\rm for} \ M_d >1\msun \\
                                  \end{array}\right. \ ,
    \ee
where $t_{\msun}$ is the time at the stage 13 of the $1\msun$ model of 
Schaller et al. 1992 (see Fig.~\ref{fig-radius}).
The applied formulae may not be very accurate, but
the qualitative behaviour after more realistic numerical calculations
will remain the same.
The lower boundary of the period
immediately after the common envelope (line denoted by C) is set by the
condition that the main sequence star should not overfill its 
Roche-lobe at the end of the common envelope. 

\section{Common Envelope Evolution: Binding Energy and Efficiency Parameter}
\label{CE}

Since our initial separation $a_{\rm i}$ is sharply defined by the
condition of case C mass transfer, and our final separations are 
constrained by the distance that can be traversed during magnetic braking
and gravitational wave radiation, we can determine the range of 
allowed common envelope efficiencies in order to form low-mass
black hole X-ray transients.

During the common envelope phase the energy needed to expel the
hydrogen envelope of the black hole progenitor is tapped from
the drop in binary orbital potential energy with efficiency 
$\alpha_{\rm ce}$ (also denoted $\eta$ in the literature):
    \be
    \frac{G M_pM_e}{\lambda R} 
    = \frac{G M_pM_e}{\lambda r_L a_i} 
    = \alpha_{ce} \left(\frac{G M_{He} M_d}{2 a_f}-\frac{G M_p M_d}{2a_i}\right)
    \ee
where $M_p$ is the total mass of the BH progenitor star just before the
common envelope forms, $M_e$ is the mass of its hydrogen envelope, 
$M_{He}$ is the mass of its core, $a_{\rm i}$ and $a_{\rm f}$ is the 
initial and final separation, before and after the common envelope, 
respectively. 
$r_L$ is the dimensionless Roche-lobe radius.
Given the parameters of the system
at the start of the common envelope, the final separation is
determined by $\lambda$, describing the structure of the giant and
$\alpha_{ce}$, the efficiency of the energy conversion. In our case
the final separation is limited as shown in Fig.~\ref{fig-possib} so
the product $\lambda\alpha_{ce}$ can be constrained. 
In the literature $\lambda=0.5$ had often been used, and a high
efficiency $\alpha_{ce}>1$ was often required in order to explain
the observations. 
However, recent detailed stellar evolution calculations by
Dewi \& Tauris (2000) show that $\lambda$ can be substantially larger.
These high values of $\lambda$ solve the problem of unrealistically
high efficiencies $\alpha_{ce}$.
For a $20\msun$ star, corresponding to $16.2\msun$ at the tip of
the AGB, Tauris \& Dewi (2001) find $\lambda$-values in a large
interval: $0.1 < \lambda < 3$ depending on the exact location of
the core mass boundary and amount of internal thermodynamic energy
included.

In Fig.~2 we plotted the possible ranges of periods
after the common envelope.
We took the ZAMS mass
of the black hole progenitor to be $20\msun$, which corresponds to
$M_p\sim 16.8\msun$ in the beginning of case~{C} mass transfer, and 
assumed $M_{He} = M_{BH} = 7\msun$. For each value of $\lambda\alpha_{ce}$ 
the two lines are for the limiting initial separations $a_{\rm i}$
(see Sect. 2).
Donor stars with masses above $1.5\msun$ can only be formed if the
common envelope efficiency $\lambda\alpha_{ce}$ is around 0.2.

\section{Life Time of X-ray Transient Sources}
\label{evol_RL}

Brown, Lee \& Bethe (1999), assuming a mass-loss rate of $10^{-9}
\msun$ yr$^{-1}$, obtained a lifetime of the X-ray transients of
$10^9$ yrs . However, as Ergma \& Fedorova (1998) discussed, the 
mass-loss rate changes as a function of time.

\begin{figure}[t!]
\begin{center}
\epsfig{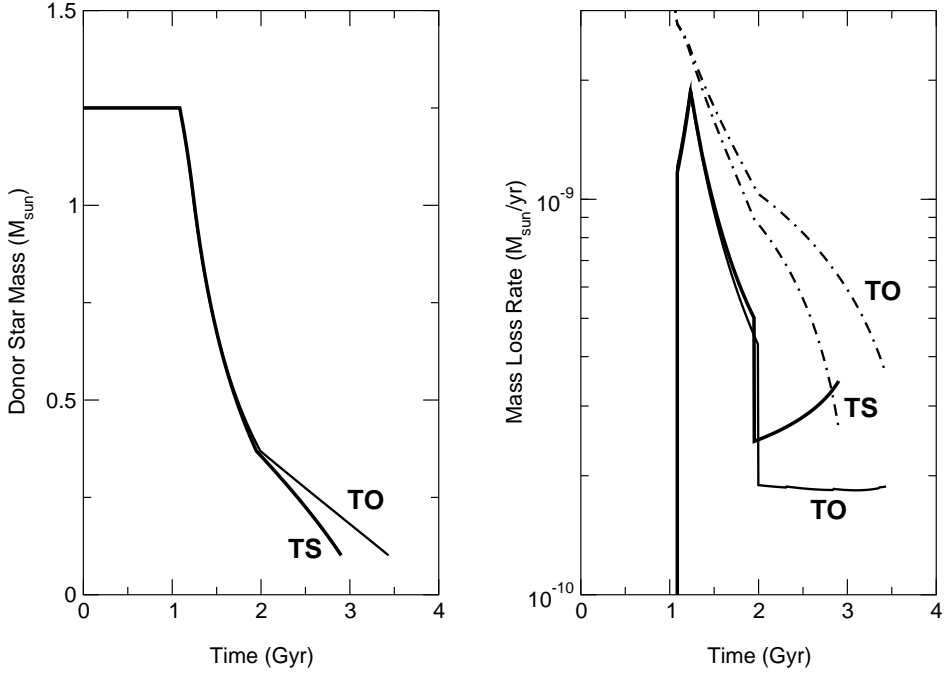}
\end{center}
\caption{Time evolution of the donor star mass and mass loss rate for
$M_{d,i}=1.25\msun$ with a $7\msun$ black hole and an initial period of 
20 hours. The dot-dashed lines on the right panel correspond to
the critical mass loss rate for the steady X-ray sources, 
eq.~(\ref{dM_crit}).
The two tracks correspond to the different time scales of
stripped donor stars (see the discussion of Sec.~\ref{evol_RL}).
}
\label{fig-mass}
\end{figure}

Once the Roche lobe overflow starts,
at any given time, the orbital separation is self determined
in order for the companion star to fill its Roche lobe,
  \be
  a = R_{d}/r_L(M_d,M_{BH})
  \ee
Hence, the mass loss rate is automatically determined by the 
feedback effects between
the orbital widening due to the mass loss and the orbital
contraction due to the gravitational wave radiation and the
magnetic braking, 
eq.~(\ref{evo_orb}).
In the numerical simulation of mass transfer due to Roche lobe overflow, 
for a given $\delta M_d$, we can get the
$\delta t$, equivalently $\dot M_d$,
by requiring the donor star fill its Roche lobe.

In Fig.~\ref{fig-mass} we plot the evolution of the donor star mass
and the mass loss rate as a function of time for $M_{d,i}=1.25\msun$ with
a $7\msun$ black hole.  There are two tracks after Roche lobe overflow.
\bi
  \item Case TO: 
          we assumed that the stripped star after Roche lobe overflow
          continues to follow exactly the
          same evolution time scale as the original star. 
  \item Case TS: 
          we assumed that the stripped donor star loses
          all its initial information.
          When the star become smaller in mass, it follows the
          time scale of the main sequence star of that reduced mass. 
  \ei
We believe the realistic situation to lie between case TO and TS. 
The radius of the donor star, which essentially determines the orbital
separation, is smaller in case TS because the donor star is
less evolved than that in TO case. 
Even though, for any given donor star mass $M_d$,
the lifetime of the donor star itself is longer in TS case, 
the smaller radius of the donor star gives the shorter 
orbital separation with
larger mass loss rates, finally decreasing the lifetime of X-ray
transients in case TS.

In
Fig.~\ref{fig-mass}, the time scales of X-ray transients are $\sim
2\times 10^9$ yrs for a $1.25\msun$ donor star. The mass of the donor
star (initially a $1.25 \msun$ star) drops quickly in the beginning with
a high mass loss rate $>10^{-9} \msun$ yr$^{-1}$, which slows down
afterwards. Any donor stars $>0.5 \msun$ show the similar
behaviour. Hence, independently of initial donor masses, most of the
time X-ray transients will appear as low-mass X-ray transients (i.e.
with a donor star mass $< 0.5 \msun$) having a mass-loss rate of $\sim
10^{-10}\msun$ yr$^{-1}$, which is consistent with the empirical rate
of $1.3\times 10^{-10}\msun$ yr$^{-1}$ found by Van~Paradijs~(1996). 
The dot-dashed lines on the right panel of Fig.~\ref{fig-mass} correspond to
the critical mass loss rate for the steady X-ray sources (King et al. 1997),
   \be
   \dot M_{crit} \approx 2.86\times 10^{-11}\
        M_{BH}^{5/6} M_d^{-1/6} P_{hrs}^{4/3}\ \msun \ {\rm yr}^{-1}.
   \label{dM_crit}
   \ee

We interpret the different masses of the main sequence companion as
resulting firstly from varying initial masses, and secondly from the
star being observed at different times in its evolution.

\section{Population Synthesis}
\label{pop}

We calculated the total number of expected X-ray transients evolving
through case~{C} mass transfer in the same way as Brown, Lee \& Bethe (1999).
From the discussions in Sec.~\ref{sec-caseC}, for a ZAMS star of mass 
$20\msun$, we see that only the interval of $235\rsun$ between
$1370\rsun$ and $1605\rsun$ of $a(t=0)$ is available for case~{C} mass
transfer. 
It is this small fractional interval that allows us to obtain a narrow
interval of values for $\lambda\alpha_{ce}$.
This is a logarithmic interval of only $\ln
(1605/1370)=0.16$ compared with our total logarithmic interval of $\ln
(4\times 10^9 \ {\rm km}/4\times 10^6\ {\rm km})=7$ (Brown, Lee, \&
Bethe 1999)
so that we have a
fraction of only $0.023$ (as compared with $0.11$ in Brown, Lee, \&
Bethe 1999).

Since the possible ZAMS range for case~{C} mass transfer is very narrow
near $20\msun$, we somewhat arbitrarily choose a ZAMS mass interval of
$19\msun < M < 22\msun$. This means (Bethe \& Brown 1998) that the
fraction of binaries with primaries in this range (assuming an IMF,
$P(m) \propto M^{-2.5}$) is: $(1.9)^{-3/2}-(2.2)^{-3/2} =0.08$
which is smaller by 
a factor of $2.5$ from that for the interval of $20-35\msun$
chosen by Brown, Lee \& Bethe (1999).

From the discussion in Sec.~\ref{evol_binary} and \ref{CE}, the upper
limit of the mass of the donor star is sensitive to the efficiency
parameter $\lambda\alpha_{ce}$. For $\lambda\alpha_{ce}=0.4$ this upper limit is
1.6 $\msun$. Assuming a
flat $q$ distribution, we get $\Delta q \approx 1/20$.  From
Sec.~\ref{evol_RL}, we take the average lifetime of X-ray transients
as $2\times 10^9$ yrs, a factor two higher than the assumed lifetime
in Brown, Lee \& Bethe (1999).

By taking the supernova rate as $2\times 10^{-2}$ yr$^{-1}$ per Galaxy,
we have the number of X-ray transients in our Galaxy as 
  \be 
  2\times 10^{-2}\ {\rm yr^{-1}}\times\frac{1}{20}\times \frac 12 \times
  0.08\times 0.023 \times 2\times 10^9 \ {\rm yr} \approx 1840, 
  \label{eq9}
  \ee
where a binarity of a $1/2$ is considered as in Brown, Lee \& Bethe
(1999).  The total numbers in the Galaxy of such systems is estimated
to be between a few hundred and a few thousand (Ergma \& Fedorova
1998).\footnote{ Of the Wijers (1996) lower limit of 3000 transient
  black hole sources, 6 out of 9 sources had a main sequence companion 
  with a short period, so he would have had $\sim 2000$ of the latter.}  
Our estimate is thus consistent with theirs.

Our birth rate for the black hole binaries obtained from Eq.~(\ref{eq9})
is $10^{-6}$ yr$^{-1}$, roughly half the Portegies Zwart et al. rate
for binaries with a neutron star.

\section{Black Hole X-ray Binaries with (sub)Giant Donors}
\label{secnova}

Transient black hole binaries with (sub)giant donors might be expected
to have followed the same scenario, but with initially higher 
companion masses and larger separations after the common envelope,
so that they start mass transfer only when the donor has evolved off
the main sequence. However, the large space velocity of 
\astrobjNovaSco, 
best explained with a large amount of mass loss (5--10$\msun$)
during the explosion in which the black hole was formed (Nelemans et
al.~1999), and the high mass of the black hole in 
\astrobjV404Cyg 
($\sim 10\msun$) suggest that the helium cores of stars around 20 $\msun$ may
not be massive enough to explain these systems.
Hence, the progenitor masses of the black hole may have been larger.


Brown \& Lee (2001) require a He core mass of $\sim 11\msun$ corresponding
to ZAMS $\sim 35\msun$ for the black hole progenitor of Nova Scorpii and
a somewhat higher He core mass for the black hole progenitor of V4641 Sgr.
The wind losses employed by Schaller et al. (1992) must be substantially
reduced if these SXTs are to be evolved in Case C mass transfer.
This would take us back to the interval of ZAMS masses
$20-35\msun$ suggested by Brown, Lee, \& Bethe (1999), possibly
even up to $\sim 40\msun$.

\section{Conclusion}

  We evolve the low-mass black-hole X-ray binaries, which are identified
  as the observed soft X-ray transients,
  and show that these systems are formed via case~{C}
  mass transfer following helium core burning phase
  of the black hole progenitor.
  Although this scenario requires a set of fine tuned conditions, we are
  able to produce a current Galactic population of $\sim$2000 objects,
  in agreement with estimates based on observations. 
  Combining the narrow interval in initial separations leading 
  to case~{C} mass transfer with the allowed narrow range of separations after 
  the common envelope evolution, we put constraints
  on the common envelope efficiency as $\lambda\alpha_{ce} \approx 0.2-0.5$.

Since our analysis requires fine tuned parameter space, more detailed
calculations of the stellar evolution of stars with ZAMS masses around $20\msun$
are required. More uncertain is the common envelope efficiency, 
which is essential for the formation of final short orbital period of
the binaries.

\section*{Acknowledgments}

This work was initiated by Gijs Nelemans' visit to Stony Brook.
We acknowledge his helpful comments and discussions.
GEB \& CHL were supported by the U.S. Department of Energy under grant
DE-FG02-88ER40388. 



\begin{thebibliography}{}
\bibitem{v4641}
   Brown, G.E. and Lee, C.-H. 2001, in preparation.
\bibitem{transient}
   Brown, G.E., Lee, C.-H., and Bethe, H.A. 1999, New Astronomy 4, 313.
\bibitem{bhllwb}
   Brown, G.E., Heger, A., Langer, N., Lee, C.-H., Wellstein, S.,
   and Bethe, H.A. 2001, astro-ph/0102379.
\bibitem{israelian}
   Brown, G.E., C.-H. Lee, Wijers, R.A.M.J., Lee, H.K., Israelian, G.,
   and Bethe, H.A. 2000, New Astronomy, 5, 191.
\bibitem{bww}
   Brown, G.E., Weingartner, J.C. and Wijers, R.A.M.J. 1996,
   ApJ, 463, 297.
\bibitem{dekool}
   De Kool, M., Van den Heuvel, E.P.J., and Pylyser, E. 1987, A\&A, 183, 47.
\bibitem{Dewi}
   Dewi, J.D.M. and Tauris, T.M. 2000, A\&A, 360, 1043.
\bibitem{ergma}
   Ergma, E. and Fedorova, A. 1998, A\&A, 338, 69.
\bibitem{ergma2}
   Ergma, E. and van den Heuvel, E.P.J. 1998, 331, L29. 
\bibitem{kalogera}
   Kalogera, V. 1999, ApJ, 521, 723.
\bibitem{kalogerawebb}
   Kalogera, V. and Webbink, R.F. 1998, ApJ, 493, 351.
\bibitem{King}
   King, A.R., Kolb, U., and Szuszkiewicz, E. 1997, ApJ, 488, 89.
\bibitem{kunz}
   Kunz, R., Jaeger, M., Mayer, A., Hammer, J.W., Staudt, G.,
   Harissopulos, S., and Paradellis, T. 2001, Phys. Rev. Lett. 86, 3244.
\bibitem{nelemans}
   Nelemans, G., Tauris, T.M., and van den Heuvel, E.P.J.  1999, 
   A\&A, 352, L87.
\bibitem{Pacz}
   Paczy\'nski, B. 1976, in P.P. Eggleton, S. Mitton, J. Whelan (eds.), 
   {\it Structure and evolution of close binary systems}, p. 75, 
   Kluwer, Dordrecht.
\bibitem{simon}
   Portegies Zwart, S., Verbunt, F., and Ergma, E. 1997,
   A\&A, 321, 207.
\bibitem{wr22}
   Rauw, G., Vreux, J.-M., Gosset, E., Hutsem\'ekers, D., Magain, P.,
   and Rochowicz, K. 1996, A\&A, 306, 771.
\bibitem{schaller}
   Schaller, G., Schaerer, D., Meynet, G., and Maeder A. 1992, 
   A\&AS, 96, 269.
\bibitem{Skumanich}
   Skumanich, A 1972, ApJ, 171, 565.
\bibitem{Stepien}
   Stepien, K. 1995, MNRAS, 274, 1019.
\bibitem{td}
   Tauris, T.M., and Dewi, J.D.M., 2001 A\&A in press.
\bibitem{paradijs}
   Van Paradijs, J. 1996, ApJ, 464, L139.
\bibitem{vf}
   Verbunt, F., in: Neutron Stars and Their Birth Events. Ed. W. Kundt,
   Kluwer, Dordrecht p.179
\bibitem{wels}
   Wellstein, S., and Langer, N. 1999, A\&A, 350, 148.
\bibitem{wijers}
   Wijers, R.A.M.J. 1996, Evolutionary Processes in Binary Stars, Vol. 327, 
   edited by Wijers et al.,
   Kluwer Acad. Publ., Dordrecht.
\bibitem{woosley}
   Woosley, S.E., and Weaver, T.A. 1995, ApJS, 101, 181.
\end{thebibliography}
\end{document}